\documentclass{aa}
\usepackage{graphicx}
\usepackage{natbib}
\bibpunct{(}{)}{;}{a}{}{,}
\begin{document}

   \title{The influence of clumping on surface brightness fits \\
	  of edge-on spiral galaxies}
 
   \author{
       A. \,Misiriotis\inst{1,2} 
        \and 
       S. Bianchi\inst{3,4,5,6}
   }
 
   \offprints{sbianchi@arcetri.astro.it}
 
   \institute{ 
   University of Crete, Physics Department, P.O. Box 2208, 710 03
   Heraklion, Creete, Greece
   \and
   Observatoire de Marseille, 2 place Le Verrier, 13248 Marseille Cedex
   4, France
   \and
   Max-Planck-Institute f\"ur Astronomie, K\"onigstuhl 17, D-69117  
   Heidelberg, Germany
   \and 
   European Southern Observatory, Karl-Schwarzschild-Strasse 2, D-85748 
   Garching, Germany
   \and 
   Max-Planck-Institute f\"ur Astrophysik, Karl-Schwarzschild-Strasse 1,
   D-85741 Garching, Germany
   \and 
   Istituto di Radioastronomia/CNR, Largo E. Fermi 5, I-50125 Firenze,
   Italy
   } 
 
   \date{Received  / Accepted } 
 
\abstract{
We have used a Monte Carlo radiative transfer code to produce 
edge-on images of dusty galactic disks, allowing a fraction of the dust to be
distributed in clumps. Synthetic images of edge-on galaxies have been 
constructed for different amounts of dust, distributions of clumps and 
fractions of dust in clumps, following the formalism of \citet{BianchiSub1999}. 
We have also considered models with stellar emission embedded in the
clumps. The synthetic images have been fitted with analytical models made 
with smooth distributions of dust, adopting the procedure developed by
\citet{XilourisSub1998} to fit optical images of real edge-on galaxies.
We have compared the parameters determined by the fit with the input
parameters of the models. For the clumping distributions adopted in this
paper, the neglect of clumping results in underestimating
the amount of dust in a galaxy. However, the underestimation is never 
larger than 40\%.
}

\maketitle

\section{Introduction}

Optical extinction by interstellar grains complicates the study of the 
stellar and dust content of a spiral galaxy. For moderate to high
optical depths, information on the intrinsic properties of each
of these components is entangled in the observed surface brightness
distribution of the galaxy. The usual way of deriving these properties
is by comparing the observations with radiative transfer calculations
in a dusty galactic model. A certain degree of complexity is needed in the 
description of the geometric distributions of a galactic model. Radiative 
transfer models with simple geometries have been proven to provide equivocal 
results \citep*{DisneyMNRAS1989}. Furthermore, scattering must be taken 
into account, given the high albedo observed for Galactic dust 
\citep*{GordonApJ1997}.

A radiative transfer code using the Monte Carlo (MC) technique can, in principle, 
accommodate any stellar and dust distribution and provide an accurate
calculation of the surface brightness for a given model
\citep*{WittApJ1992,BianchiApJ1996,DeJongA&A1996b}. However, the MC 
method is time consuming and thus can not be easily
included in fitting procedures. For such procedures an approximate treatment
of the radiative transfer is more appropriate.

The first approach of such a procedure was introduced by 
Kylafis \& Bahcall (1987, hereafter KB),
where vertical profiles of
a model galaxy were fitted to observations of NGC~891. In KB's work,
scattering is calculated up to the first
order and an approximation is introduced for the higher order contributions
while the distribution of the dust is assumed to be smooth and exponential
along the radial and vertical directions.
A similar approximation for the treatment of scattering is adopted by
Silva et al. (1998).
Other works prefer an exact solution for scattering within simple 
geometries \citep*{BruzualApJ1988,CorradiMNRAS1996,XuA&A1995}. 

Using the KB method, the radiative transfer equation
can be solved for a wide variety of
geometries \citep{ByunApJ1994}. Xilouris et al. (1997;1998)
improved the original idea of KB and implemented 
a technique to fit the surface brightness distribution of edge-on spirals.
In edge-on galaxies, the effects of extinction are maximized
and it is possible to  separate the stellar and dust components.
The method has been successfully applied to a sample of 
seven edge-on galaxies \citep{XilourisSub1998} concluding 
that late-type spiral galaxies have a moderate opacity, with a mean face-on optical 
depth in the B-band $\tau_\mathrm{B}\approx 0.8$.
As in most of the models used for the description of spiral galaxies,
\citet{XilourisSub1998} adopt smooth exponential distributions to 
describe the stellar and dust disks.

Real galaxies exhibit a wide 
variety of inhomogeneities, like spiral arms, bars, clumps. While it 
is possible to include these structures in the solution, when fitting an 
observed image it is desirable to deal with the simplest possible description, 
in order to limit the number of model parameters. Complex models can then
be used to
test the reliability of the description obtained with the simple models.
This is done, for example, in \citet{MisiriotisA&A2000},
where synthetic edge-on images of model galaxies with a spiral structure 
are fitted with a smooth exponential model.
They find that plain exponential distributions provide a good 
description of the galactic disks. 
The derivation of the stellar and dust parameters is only slightly affected by 
the spiral pattern.

In this paper we perform a similar exercise to study the influence of
dust clumping on the fit of edge-on galaxies. In most cases, a clumpy medium 
exhibits higher transparency than a smooth one of the same dust mass. 
Therefore, the comparison of real images with smooth models may result in 
an underestimation of the dust content in a galaxy. Here we use the MC 
method  presented by \citet[][hereafter BFDA]{BianchiSub1999} to create 
images of clumpy galactic models. For the parametrisation of BFDA, the
maximum difference between clumpy and homogeneous models can be seen in
the edge-on case. Monte Carlo models of highly inclined galaxies
had been produced also by \citet{KuchinskiAJ1998}, with a different
parametrisation for the clump distribution. With respect to their result,
the influence of clumping in the edge-on case is larger for the model
adopted here (see the discussion in BFDA). 
The MC images of BFDA will be fitted with the KB 
model as in \citet{MisiriotisA&A2000}, aiming at determining the efficiency of 
clumping in hiding the dust when a galaxy is seen edge-on. 

Sect.~\ref{fits} describes the method adopted for the comparison and the 
MC models we have used to produce the synthetic images. A model 
with exponentially distributed clumps developed for this work is presented 
in the Appendix.  The results are presented in 
Sect.~\ref{results} and a  summary is given in Sect.~\ref{conclu}.

\section{Our Method}
\label{fits}

\begin {figure*}[!ht]
\sidecaption
\includegraphics[width=12cm]{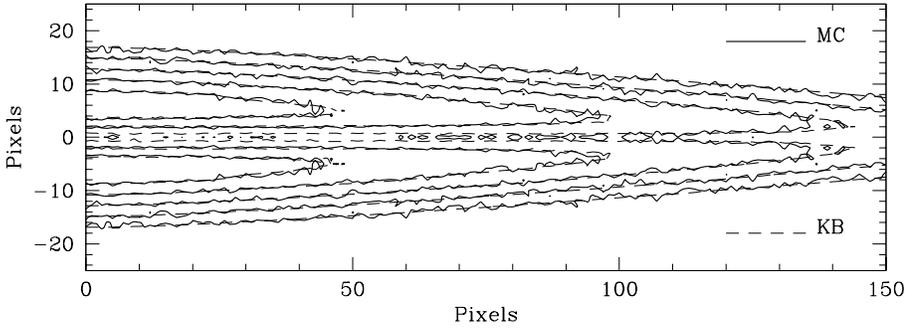}
\caption{Contour plot of a synthetic edge-on image from a Monte Carlo 
smooth model with $\tau_\mathrm{V}$=1. Overplotted are isophotes from a 
\citet{KylafisApJ1987} type model.}
\label{fit_iso}
\end{figure*}

Following the procedure described by \citet{MisiriotisA&A2000}, we 
fit  images made with the KB method to the MC images of a clumpy model galaxy.
Both the MC and the KB models presented here are computed for the V-band. 
The \citet{HenyeyApJ1941} phase function is used with the scattering parameters
in the V band
albedo $\omega=0.6$ and asymmetry $g=0.6$ \citep{GordonApJ1997}.

The parameters defining the smooth models are
the central face-on surface brightness $I_\star$,
the radial scalelength $\alpha_\star$ and
the vertical scalelength $\beta_\star$ of the stellar disk, 
the V-band central face-on optical depth $\tau_\mathrm{V}$, 
the radial scalelength $\alpha_d$ and 
the vertical scalelength $\beta_d$ of the dust disk. 
The total intrinsic luminosity $L_\star$ is equal to
$4 \pi I_\star \alpha_\star^2 $ while the dust mass $M_d$
is propotional to $\tau_\mathrm{V} \alpha_d^2$.
The results from the fits are compared to the input parameters 
of the MC images and are analyzed as a function of the fraction of 
dust locked into clumps $f_\mathrm{c}$. Additionally, 
we study the importance of scattering in the fitted models.

The disk parameters adopted for the MC models are those
used in BFDA. For the stellar disk, we have adopted a ratio between
radial and vertical scalelengths $\alpha_\star/\beta_\star=11.5$. The 
smooth dust disk has the same radial scalelength as the stellar one and
a vertical scalelength $\beta_\mathrm{d}=0.4 \beta_\star$. In both the 
simulation and the fitting procedure the dust distribution has
been truncated at 4.6  radial scalelengths along the radial 
direction and at 4 vertical scalelengths
along the vertical direction.

In the clumpy images,
$\tau_\mathrm{V}$ defines the total mass of dust in the model, i.e. 
for the same $\tau_\mathrm{V}$ the clumpy and the smooth models 
have the same dust mass. In partially clumpy models,
a fraction $f_\mathrm{c}$ of the total dust mass is distributed in clumps, while the
remaining is distributed in a smooth disk of central face-on optical
depth $\tau_\mathrm{V} (1-f_\mathrm{c})$. The number of the clumps $N_c$
 is determined as described in BFDA.

To cover the full range of $\tau_\mathrm{V}$ and $f_\mathrm{c}$ we considered
the following optical depths :
$\tau_\mathrm{V}$=0.5, 1, 2 and 5. For each optical depth we produced images
with $f_\mathrm{c}$= 0 (smooth case), 0.25, 0.5, 0.75 and 1.0, resulting in a set of
20 images. Several different sets can be produced depending on how
we distribute the dust clumps. We have decided to produce 3 sets.

In the set named ``molecular'', the dust clumps
follow the same distribution as the molecular gas in the
Galaxy and their characteristics are based on the Giant Molecular Clouds.
Details on these models can be found in BFDA. In this set,
 the $f_\mathrm{c}$=1 case
is not considered because it would mean that all the dust is distributed in a
molecular ring and an exponential model is not reccommended
to reproduce such a distribution.

A second set of models named ``exponential'' was produced with the
dust clumps following the same exponential distribution as the smooth dust disk
while the stars were smoothly distributed.  
The exponential models are described in detail
in the Appendix.

In the third set of models named ``embedded'' the clumps
are distributed exponentially, while
a fraction of starlight $f_\mathrm{emb}$=0.5 was 
embedded in the dust clumps. 
For the case of $f_\mathrm{c}$=0, there can be no $f_{\mathrm{emb}}$=0.5 
because when there are no clumps there is nowhere to embed
the embedded starlight.

The MC images of the clumpy model galaxies were constructed to be as similar
as possible to observations.  The images cover 151 pixels along the major axis 
from the galactic center to the border of the image, while the radial scalelength
of the stellar and the dust disk is equal to 50 pixels.
For each image, photons are gathered in a
finite inclination band \citep{BianchiApJ1996}. which was chosen to be
1$^\circ$. The number of photons for the MC
simulations was chosen so that a realistic signal to noise ratio was
achieved \citep{BianchiApJ1996}.

The fitting was performed as in \citet{MisiriotisA&A2000}, by minimizing the 
sum of the squares of the differences between the MC and the KB images.
In Fig.~\ref{fit_iso} we present a contour plot of an edge-on 
image from a MC smooth model ($f_\mathrm{c}$=0)
together with the isophotes of the same model image, produced using the
KB method. The coincidence of the 
MC and KB model images demonstrates the consistency of the two methods.

\section{Results of the fits}
\label{results}

\subsection{Testing of the codes and the effects of scattering}
Since the two radiative transfer methods (BFDA and KB)
used in this work have been 
developed independently and with different algorithms, we checked for 
their consistency in several ways. We first produced images
without scattering and without clumps using the MC method and fitted them
with the KB method. The fits were able to retrieve all the
parameters with a scatter smaller than 4\%.

We then analyzed the smooth models ($f_\mathrm{c}=0$) described in the previous
section. Two fits were produced for each MC image. In
the first, the fitting model is computed without scattering, while in the
second scattering is taken into account. This way we can quantify both
the importance of scattering in fits of edge-on galaxies and
the approximation of KB in their treatment of scattering.

\begin {figure}[b]
\resizebox{\hsize}{!}{\includegraphics{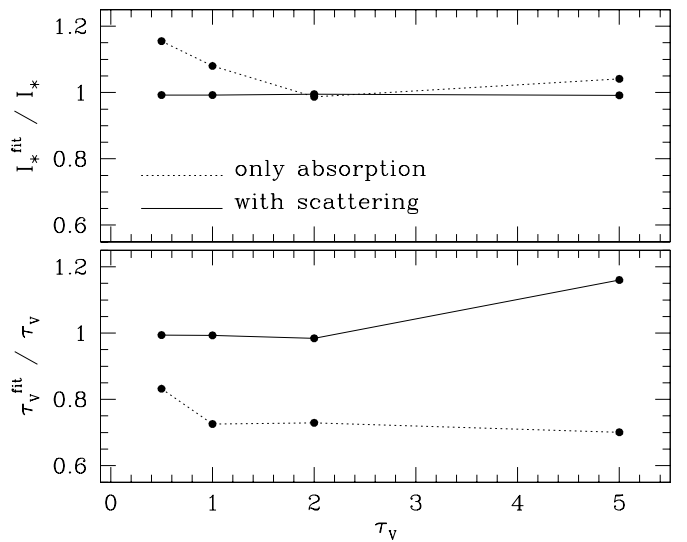}}
\caption{The derived $I_\star$ and $\tau_\mathrm{V}$ normalized to the
input values as functions of the optical depth of the model. Results
for fits with and without scattering are shown.}
\label{smooth}
\end{figure}

In Fig.~\ref{smooth} we show the parameters $I_\star$ and $\tau_\mathrm{V}$
whose behavior changes the
most when passing from an absorption-only fit to a fit including
scattering. It is well-known that
scattering reduces the effective absorption of a dusty medium, thus
resulting in an increased surface brightness.
When scattering is not included 
(dotted lines in Fig.~\ref{smooth}), the fitting
procedure compensates for this by reducing the optical depth of
the dust disk and by increasing the
central surface brightness of the stellar disk. 
The underestimate of the dust content is the dominant effect, because
the stellar luminosity is significantly constrained by the parts of the 
galaxy that lie outside of the dust lane, 
where the absorption is negligible.
Absorption-only fits result in an
underestimate of the face-on optical depth
ranging from 10\% for $\tau_\mathrm{V}$=0.5 to 25\% for $\tau_\mathrm{V}$=5. 
As for the central face-on surface brightness ($I_\star$), 
the maximum discrepancy is an overestimation by 10\% for $\tau_\mathrm{V}$=0.5 

When scattering is included, the fitted parameters are very close to the 
input values (solid lines in Fig.~\ref{smooth}). The $\tau_\mathrm{V}$ plot 
confirms the validity of the approximation for scattering introduced by
\citet{KylafisApJ1987}, at least for low-to-moderate optical depths.
Even for $\tau_\mathrm{V}$=5, the derived optical
depth is only 15\% larger than the input value. All the other parameters derived 
from fits with scattering were found to vary within less than 
4\% of the input values.

\subsection{The effect of clumping}
\begin {figure*}[!t]
\centering
\includegraphics[width=17cm]{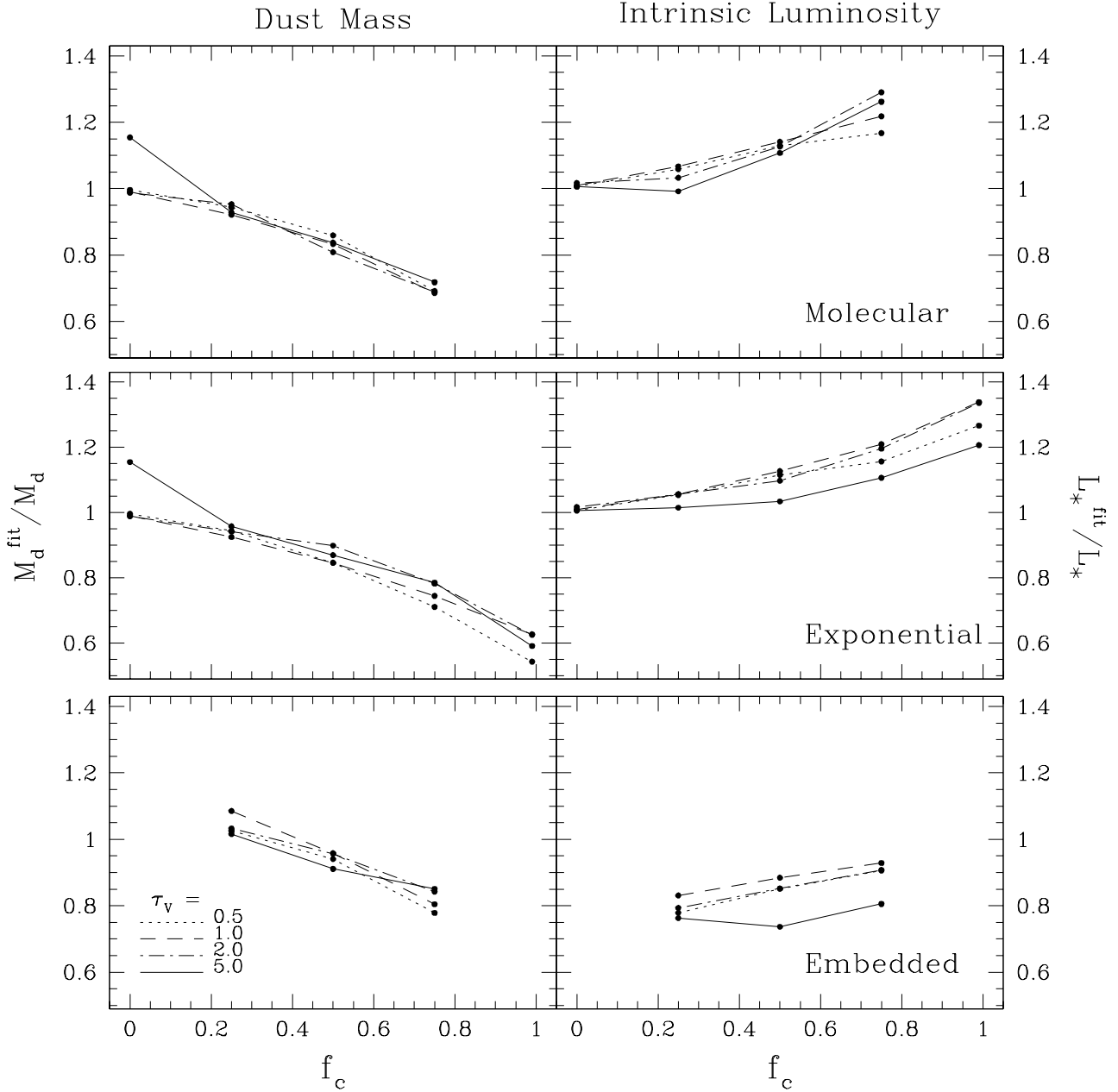}
\caption{Normalized dust mass (left column) and normalized
intrinsic luminosity (right column)
derived by fitting the Monte Carlo simulations of Section~\ref{fits}, as
functions of the fraction $f_\mathrm{c}$ of dust locked up in clumps. 
The values 
have been normalized to the input dust mass and intrinsic luminosity. Results
are presented for models with optical depths $\tau_\mathrm{V}$= 0.5
(dotted lines), 1 (dashed), 2 (dot-dashed) and 5 (solid). The plots in 
the top row show the results for the clumpy model of BFDA with fraction of
embedded stellar emission $f_\mathrm{emb}$=0; the middle row refers to the
models with clumps distributed exponentially and $f_\mathrm{emb}$=0; the
bottom row to exponentially distributed clumps, 
but with $f_\mathrm{emb}$=0.5. Results 
for the smooth models ($f_\mathrm{c}=0$) are plotted in both the 
top and middle row .}
\label{fig_res}
\end{figure*}

The quality of the fits is poorer when synthetic images from clumpy models are
analyzed. The sum of the squares of the differences between input data
and fitted models increases by about an order of magnitude. Furthermore,
in many cases it was possible to find a few equally acceptable fits, 
corresponding to local minima.
While some parameters of KB models may change from minimum to minimum, 
we found that the derived dust mass and intrinsic luminosity are nearly
the same.

In Fig.~\ref{fig_res} we have
plotted the derived normalized dust mass and intrinsic luminosity, 
as functions 
of $f_\mathrm{c}$ and of $\tau_\mathrm{V}$, for the ``molecular'' models,
(BFDA; top row); for the ``exponential'' models
 (middle row) and for the ``embedded'' models (bottom row).

The derived dust masses, normalized to the input values, are shown in the 
left column of Fig.~\ref{fig_res}.  The trend of the dust mass with
$f_\mathrm{c}$ does not depend on the type of clumping or on the optical depth 
of the model, the amount of detected dust decreases with
increasing $f_\mathrm{c}$. 
For the two sets of models
without embedded stellar emission (``molecular'' and ``exponential''),
the dust mass is underestimated by
less than 10\% for $f_\mathrm{c}$=0.25 to about 20-30\% for $f_\mathrm{c}$=0.75.
For the ``exponential'' set for $f_\mathrm{c}$=1,
the underestimate reaches values of 40\%. For the ``embedded'' models
the fitting procedure recovers the input dust
quite well. All the dust is detected for $f_\mathrm{c}$=0.25 and only 20\% is missed for $f_\mathrm{c}$=0.75.

The results of the fitting procedure
for the intrinsic, dust-free, luminosity $L_\star$ are shown in
the right column of Fig.~\ref{fig_res}, again normalized to the input
value. As for the dust mass, the trends with $f_\mathrm{c}$ for 
different models are very similar and do not depend 
sensitively on the value of $\tau_\mathrm{V}$.
The fitting model overestimates the luminosity with increasing
$f_\mathrm{c}$. For $f_\mathrm{c}$=0.75 and $f_\mathrm{emb}=0$, the derived luminosity is about 20\% larger than the input value. 

The overestimate of the luminosity drives the fitting procedure
to increase the dust mass in order to obscure the additional
starlight. If the luminosity was not overestimated then even less
dust would be needed to fit the synthetic images.
A simple major-axis analysis, i.e. a comparison of the radial profiles
along the dust lane suggests that the smooth model detects only 
half of the dust mass (BFDA) and leads to the conclusion that
clumping causes an underestimate of the face-on optical depth.

This is not the case for the fits shown here. The fitting procedure
tries to interpret the reduced extinction caused by clumping by
decreasing the dust radial scalelength $\alpha_d$ and keeping the
optical depth high (actually, to a value similar to the input
one), rather than decreasing $\tau_\mathrm{V}$ and keeping $\alpha_d$
fixed. The reason for this may be that the surface brightness in 
models with clumping
departs from that of a smooth exponential disk, when {\em all} the 
image, and not only a single profile, is fitted. 
Given the general trend of the retrieved data, it does not seem probable
that we have systematically hit local minima of the fit. 

We have shown that the trend of dust mass underestimation is nearly
independent of the optical depth of the model. This may suggest that
the same percentage of dust is missed when fitting a galaxy at different
wavelengths. Therefore, clumping should not affect the determination of 
the extinction law from fits of a galaxy in different bands. 
As long as the dust properties do not vary significantly with
wavelength, as in the optical range, it is possible to interpret the 
absence of a trend with $\tau_\mathrm{V}$ as an absence of a trend with
$\lambda$.

\section{Summary}
\label{conclu}

We have used a Monte Carlo code for the radiative transfer in a
clumpy galactic model to produce synthetic edge-on images of spiral 
galaxies in the V-band.
The images have then been analyzed using the fitting procedure of
\citet{XilourisSub1998}, that uses the method of \citet{KylafisApJ1987} 
to compute the radiative transfer in smooth galactic models. We first
checked for the consistency of the two independent codes, by analyzing 
smooth MC images. Despite the differences in the solution of the
radiative transfer, it was possible to retrieve the parameters of
the MC models within 4\% of their original value and we confirmed
the importance of scattering in the radiative transfer simulations. 
The dust mass is significantly underestimated when scattering is not 
included in the fitting routine.

We have then studied the influence of clumping on the derivation of the
parameters, when a synthetic image of an edge-on clumpy galactic model
is analyzed through a simpler (and more controllable) smooth model. As 
expected, using a smooth model when fitting images of edge-on 
galaxies leads to an underestimation of the dust content of the galaxy. 
For the distribution of clumps adopted in this paper, about 20-30\% of 
the dust mass is missed by the fitting, the value depending more on the 
fraction of dust located in clumps rather than on the details of the clumps 
distribution or the total dust mass of the model. In addition, clumping
alters the surface brightness distribution with respect to that
of a smooth exponential model, thus making the fit more difficult. This 
may explain the overestimation of the intrinsic luminosity in models
with clumping. 

We stress again here that our work is limited to the edge-on
case, for which convergent fits can be obtained with the KB technique.
The results may also depend on the clump distribution and on the geometrical 
parameters adopted here. However, it is interesting to note that other works 
find a small influence of clumping in edge-on galaxies.
\citet{KuchinskiAJ1998} produced radiative transfer models of disk
galaxies including clumping. The distribution of clumps is different
from the one adopted here: the clumpy medium is characterised by a
costant filling factor throughout the whole disk and by a density
contrast with the homogeneous medium \citep[][ see also BFDA for a
comparison with the present models]{WittApJ1996}.
Comparing colour gradients across the extinction lane of highly inclined
galaxies with the model results, they find that the inferred opacity
of the dust disk is largely insensitive to the difference between clumpy
and homogeneous dust distributions.

Recent analyses of dust emission at $\lambda>100\mu$m suggest that the 
amount of dust in spiral galaxies is larger than what derived from fits 
of edge-on surface brightness profiles
\citep{BianchiA&A2000,PopescuA&A2000,MisiriotisA&A2001}.
Clumping has been imputed to as a possible cause for this discrepancy:
dust in star-forming regions may contribute dominantly to the Far
Infrared emission without being seen through its extinction effects.
However, the underestimation produced by the clumpy distributions 
adopted in this paper is always smaller than 40\%, unable to explain the 
magnitude of the effect. Different dust/clump distributions may be needed. 
\citet{PopescuA&A2000} and \citet{MisiriotisA&A2001} find that the long 
wavelength emission observed in two galaxies (namely, NGC 891 and NGC 5907) 
can be explained by including a second dust disk, at least as massive as 
the dust disk causing the observed extinction lane and thin enough to escape 
detection through the technique of surface brightness fitting.


\begin{acknowledgements}
We would like to thank Professor N. D. Kylafis for his suggestions and
assistance concerning this work, and the referee, Professor A. Witt, for
thoughtful comments.
\end{acknowledgements}

\appendix

\section{Models with exponentially distributed clumps}
\label{ex_models}

In BFDA it was assumed that all the material associated with 
the molecular phase in a spiral galaxy is in Giant Molecular Clouds.
Radiative transfer models were produced assuming that dust clumps are
distributed as the molecular gas in the Galaxy, i.e. in a ring-like
structure. However, most of luminous, face-on, late-type galaxies, and
a good fraction of early-type spirals, show a radial exponential
distribution for the H$_2$ column density \citep{YoungARA&A1991}.
For the sake of completeness, we briefly present here a set of models 
with an exponential distribution for the dust clumps. The clumps distribution 
has the same scalelengths as the smooth dust disk. All the other parameters 
are the same as in BFDA.

\begin {figure}[!t]
\resizebox{\hsize}{!}{\includegraphics{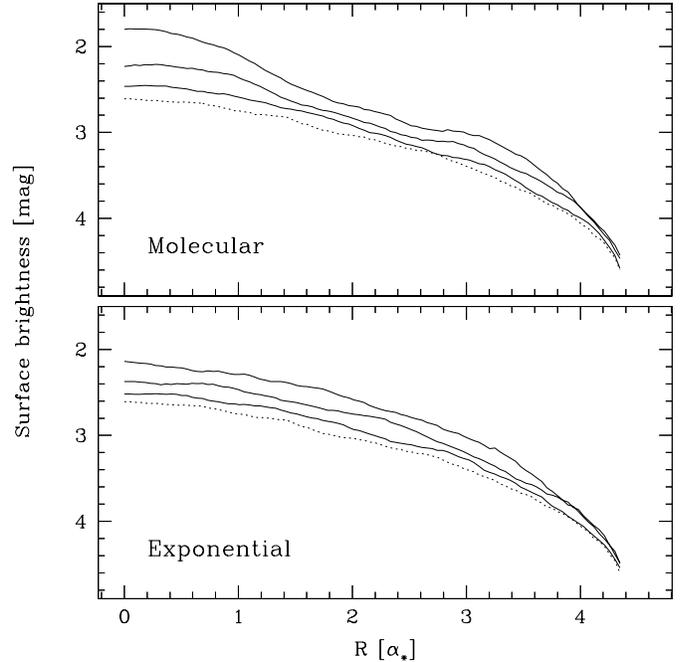}}
\caption{Major-axis edge-on profiles for $\tau_\mathrm{V}$=2 models with 
molecular (top panel) and exponential (bottom panel) clumping, for
$f_\mathrm{c}$=0.25,0.5 and 0.75 (from bottom to top, solid lines). 
As a reference, we also show the profile for a $\tau_\mathrm{V}$=2 smooth 
model (dotted lines).}
\label{figclump}
\end{figure}

The effects of clumping in these models are not very different from
those obtained adopting the Galactic H$_2$ distribution. Exponentially
distributed clumps make the models less transparent, but only slightly. 
For example, in the $\tau_V$=2 model with fraction of gas in clumps
f$_c$=0.75, 15\% of the radiation is absorbed, while in BFDA it was 12\%. 
We recall that in a smooth model 19\% of the radiation is absorbed. For 
models with embedded stellar emission, the difference is even smaller, 
because of the increased contribution of the absorption within each clump. 
The number of clumps and their opacity are independent of the assumed clump
distribution and are the same as in BFDA.

As in BFDA, the largest differences between homogeneous and clumpy
simulations are for the edge-on profiles. In Fig.~\ref{figclump} we show
the edge-on major-axis profiles for the BFDA models and for the models
with exponential clumping, for $\tau_\mathrm{V}$=2,
$f_\mathrm{c}$=0.25,0.5,0.75 and f$_\mathrm{emb}$=0. The profile for a
smooth $\tau_\mathrm{V}$=2 disk is shown as a reference. 
Because of the exponential distribution, the profiles of the new clumpy
models are similar to those for smooth exponential disks, independently
of $f_\mathrm{c}$ and f$_\mathrm{emb}$. Instead, BFDA models show the
effects of the ring-like clumping distribution, even when f$_\mathrm{emb}$=0.

\end{document}